\documentclass[pre,twocolumn,groupedaddress]{revtex4}
\usepackage{graphics,epsfig,amssymb}

\newcommand{\deriv}[2]{\frac{\mathrm{d}#1}{\mathrm{d}#2}} 

\begin{document}

\title{Interplay between function and structure in complex networks} 
\author{Timothy C. Jarrett$^+$, Douglas J. Ashton$^*$, Mark Fricker$^{**}$ and Neil F. Johnson$^+$}

\affiliation{$^+$Physics Department, Oxford University, Oxford, OX1 3PU, U.K.}

\affiliation{$^*$Physics Department, University of Nottingham, Nottingham, NG7 2RD, U.K.}

\affiliation{$^{**}$Department of Plant Sciences, Oxford University, Oxford, OX1 3RB, U.K.}

\date{\today}

\begin{abstract}
We show that abrupt structural transitions can arise in functionally optimal networks, driven by small changes in the level of transport congestion. Our results offer an explanation as to why so many diverse species of network structure arise in Nature (e.g. fungal systems) under essentially the same environmental conditions. Our findings are based on an exactly solvable model system which mimics a variety of biological and social networks. 
We then extend our analysis by introducing a novel renormalization scheme involving cost motifs, to describe analytically the average shortest path across multiple-ring-and-hub networks. As a consequence, we uncover a `skin effect' whereby the structure of the inner multi-ring core can cease to play any role in terms of determining the average shortest path across the network.

\end{abstract}

\maketitle

\section{Introduction}\label{sec:intro}

There is much interest in the {\em structure} of the complex networks which are
observed throughout the natural, biological and social sciences
\cite{newman,newman2,watts98,nets,nets2,charges,central,search,gradient,prl,dm}.
The interplay between structure and function in complex networks has become a
major research topic in physics, biology, informatics, and sociology
\cite{watts98,nets,nets2,charges,central, search, gradient}. For example, the
very same links, nodes and hubs that help create short-cuts in space for
transport may become congested due to increased traffic yielding an increase in
transit time \cite{charges}. Unfortunately there are very few analytic results
available concerning network congestion and optimal pathways in real-world
networks \cite{charges,central,search,gradient,danila,moore,
gastner}.

The physics community, in particular, hopes that a certain universality might
exist among such networks. On the other hand, the biological community knows all
too well that a wide diversity of structural forms can arise under very similar
environmental conditions. In medicine, cancer tumors found growing in a given
organ can have very different vascular networks. In plant biology, branching
networks of plant roots or aerial shoots from different species can co-exist in
very similar environments, yet look remarkably different in terms of their
structure.  Mycelial fungi \cite{mark} provide a particularly good example, as
can be seen in Figs. 1(a) and 1(b) which show different species of fungi forming
networks with varying degrees of lateral connections (anastomoses).  Not only do
fungi represent a wide class of anastomosing, predominantly planar, transport
networks, but they have close parallels in other domains, including vascular
networks, road and rail transport systems, river networks and manufacturing
supply chains. But given that such biological systems could adapt their
structures over time in order to optimize their functional properties, why do we
observe such different structures as shown in Figs. 1(a)  and 1(b) under
essentially the same environmental conditions?

In this paper, we provide exact analytic results for the effects of congestion
costs in networks with a combined ring-and-star topology \cite{prl}.  We thus address the
question above by showing that quite different network structures
can indeed share very similar values of the functional characteristics relevant
to growth. We also show that small changes in the level of network
congestion can induce abrupt changes in the optimal network structure. In
addition to the theoretical interest of such phase-like structural transitions,
our results suggest that a natural diversity of network structures should arise
under essentially the same environmental conditions -- as is indeed observed for
systems such as fungi (see Figs. 1(a) and (b)). We then extend this analysis by
introducing a novel renormalization scheme involving cost motifs, to describe
analytically the average shortest path across multiple-ring-and-hub networks. 
We note that although some of the findings of Ref. \cite{central} appear similar 
to the present ones in terms of the wording of the conclusions, the context 
and structures considered are quite different -- in addition, the results in 
the present paper are analytic and are obtained in a completely different way.

As
a consequence of the present analysis, we uncover an interesting `skin effect' whereby the structure of
the inner multi-ring core can cease to play any functional role in terms of determining the
average shortest path across the network. The implication is that any food that
is found on the perimeter of the network structure, can be transported across
the structure {\em without} having to go through the central core -- and as a
result, the network structure in the central core may begin to die out because
of a lack of nutrient. Interestingly, there is experimental evidence that real
fungal networks  \cite{mark} do indeed exhibit such a skin-effect. Other
real-world examples in which an inner network core ceases to be fed by nutrients
being supplied from the perimeter, and hence dies out, include the vasculature
inside the necrotic (i.e. dead) region in a growing cancer tumour, and the inner
core of a growing coral reef.  

Our analytically-solvable model system is inspired by the transport properties
of real fungi (see Fig. 1(c)). A primary functional property of an organism such
as a fungus, is to distribute nutrients efficiently around its network structure
in order to survive. Indeed, fungi need to transport food (carbon (C), nitrogen
(N) and phosphorous (P)) efficiently from a localized source encountered on
their perimeter across the structure to other parts of the organism. In the
absence of any transport congestion effects, the average shortest path would be
through the center -- however the fungus faces the possibility of `food
congestion' in the central region since the mycelial tubes carrying the food do
not have infinite capacity.  Hence the organism must somehow `decide' how many
pathways to build to the center in order to ensure nutrients get passed across
the structure in a reasonably short time. In other words, the fungus -- either
in real-time or as a result of evolutionary forces -- chooses a particular
connectivity to the central hub. But why should different fungi (Figs. 1(a) and
(b)) choose such different solutions under essentially the same environmental
conditions? Which one corresponds to the optimal structure? Here we will show that, surprisingly, various structurally distinct fungi can each be functionally optimal at the same time.

Figure 1(d) shows our model's ring-and-hub structure. Although only mimicking
the network geometry of naturally occurring fungi (Figs. 1(a) and (b)), it is
actually a very realistic model for current experiments in both fungal and
slime-mold systems. In particular, experiments have already been carried out
with food-sources placed at peripheral nodes for fungi (Fig. 1(e)) and
slime-mold \cite{slime} with the resulting network structures showing a wide
range of distinct and complex morphologies \cite{slime}. We use the term  `hub' very
generally, to describe a central portion of the network where many paths may
pass but where significant transport delays might arise. Such delays represent
an effective cost for passing through the hub. In practice, this delay may be
due to (i) direct congestion at some central junction, or (ii) transport through
some central portion which itself has a network structure (e.g. the inner ring
of the Houston road network  in Fig. 1(f)). We return to this point later on.

\begin{figure}[ht]
\includegraphics[width=.50\textwidth]{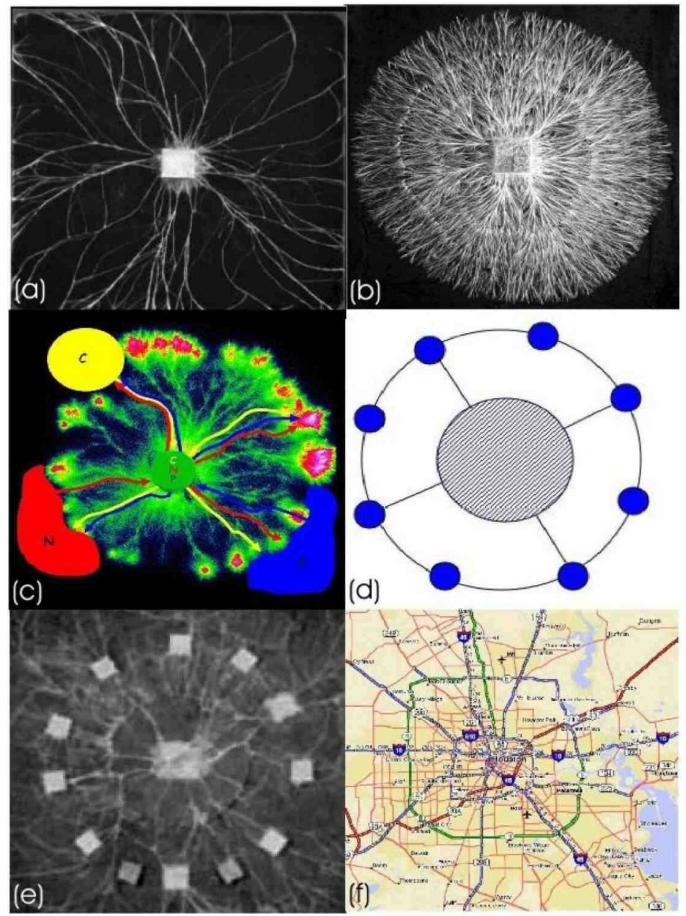}
\caption{
(a) Typical network for the fungus {\em Resinicium bicolour}.     
(b) Typical network for the fungus {\em Hypholoma fasciculare}. This network has a far denser set of  connections than (a), yet both are able to thrive in the same environmental conditions. (c) Schematic representation of the nutrient flows through a mycelial fungus. The food resources (carbon (C), nitrogen (N), phosphorous (P)) need to be transported efficiently across the entire structure.
(d) Our analytically-solvable model showing radial connections from peripheral nodes to an effective hub.
(e) Mycelial fungus {\em Phanerochaete velutina} after 98 days growing from a central hub-like resource. From day 48, the system is 
    supplied with pairs of fresh 4 cm$^3$ wood resources at 10 day intervals. The 
    resultant network has both radial and circumferential connections, as in our model (Fig. 1(d)).
(f) The man-made road network in Houston, showing a complicated inner `hub' which contains an
    embedded inner ring.}
\label{fig:figure1}
\end{figure}

\section{The Model}\label{sec:Model}

\subsection{The Dorogovtsev-Mendes model of a Small-World network}\label{subsec:DM}
We begin by introducing the Dorogovtsev-Mendes (hereon DM) model \cite{dm}
of a small world network. The DM-model consists of a ring-hub structure,
and places $n$ nodes around a ring, each connected to their nearest
neighbour with a link of unit length.  Links around the ring can
either be directed in the ``directed'' model or undirected in the
``undirected'' model. With a probability $p$ each node is
connected to the central hub by a link of length $\frac{1}{2}$, and
these links are undirected in both models.

We may proceed to solve this model, as in \cite{dm}, by first finding the
probability $P(\ell,m)$ that the shortest path between any two nodes on the ring
is $\ell$, given that they are separated around the ring by length $m$.  These
expressions can be found explicitly for both directed and undirected models.
Summing over all $m$ for a given $\ell$ and dividing by $(n-1)$ yields the
probability $P(\ell)$ that the shortest path between two randomly selected nodes
is of length $\ell$. The average value for the shortest path across the network
is then $\label{single lbar def} \bar{\ell}=\sum_{\ell=1}^{n-1}\ell P(\ell)$.
For the undirected model,  the expressions are more cumbersome due to the
additional possible paths with equal length.  However, if we define $n
P(\ell)\equiv Q(z,\rho)$ where $\rho\equiv p n$ and $z\equiv \ell/n$, a simple
relationship may be found between the undirected and directed models in the limit
$n\rightarrow \infty$ with $p \rightarrow 0$, that is
$Q_{undir}(z,\rho)=2Q_{dir}(2z,\rho)$.  Thus the ``directed'' and ``undirected''
models only differ in this limit by a factor of two:
$z \rightarrow 2z$, with $z$ now running from $0$ to $1/2$.

\subsection{The addition of congestion costs}
We generalize the DM model of section \ref{subsec:DM} to include a cost, $c$,
for passing through the central hub\cite{prl}.  This cost $c$ is expressed as an
additional path-length, however it could also be expressed as a time delay or
reduction in flow-rate for transport and supply-chain problems.  We then consider
a number of cases for the structure of such a cost, e.g. a constant cost $c$ where $c$ 
is independent of how many
connections the hub already has, i.e.  $c$ is independent of how `busy' the hub
is; a linear cost $c$ where $c$ grows linearly  with the number of connections
to the hub, and hence varies as $\rho \equiv np$; or nonlinear cost $c$ where
$c$ grows according to a number of nonlinear cost-functions.

For a general, non-zero cost $c$ that is independent of $\ell$ and $m$, we can
write (for a network with directed links):
{\small \begin{eqnarray} P(\ell ,\ell \leq c) &=& \frac{1}{n-1} \\ P(\ell < m, \ell > c) &=&
      (\ell-c)p^2(1-p)^{\ell-c-1} \label{p-ell-lt-m} \\ P(\ell = m, \ell > c) &=&
      1-p^2\sum_{i-c=1}^{\ell-c-1}(i-c)(1-p)^{(i-c)-1} \end{eqnarray}} \noindent
      Performing the summation gives: \begin{equation} P(\ell = m, \ell > c) =
      (1+(\ell-c-1)p)(1-p)^{\ell-c-1} \end{equation} The shortest path distribution is
      hence: 
{\small \begin{displaymath} 
P(\ell) = \left\{ \begin{array}{ll}
\frac{1}{n-1} & \ \ \textrm{$\forall$ $\ell \leq c$}\\
\frac{1}{n-1}\bigl[1+(\ell -c-1) p \\
\ \ \ \ \ \ + (n-1-\ell )(\ell -c)p^2 \bigr](1-p)^{\ell-c-1} & 
\ \ \textrm{$\forall$  $\ell > c$}
\end{array} \right.  
\end{displaymath}}
Using the same analysis for undirected links yields a simple relationship
between the directed and undirected models.  Introducing the variable $\gamma
\equiv \frac{c}{n}$ with $z$ and $\rho$ as before, we may define $nP(\ell)
  \equiv Q(z, \gamma, \rho)$ and hence find in the limit $p \rightarrow 0$, $n
  \rightarrow \infty$ that $Q_{undir}(z,\gamma,\rho)=2Q_{dir}(2z,
      2\gamma,\rho)$.  For a fixed cost, not dependent on network size or the
  connectivity, this analysis is straightforward.  Paths of length $l \leq c$
  are prevented from using the central hub, while for $l>c$ the distribution
  $P(l)$ is similar to that of Ref. \cite{dm}.

\begin{figure}[ht]
\includegraphics[width=.35\textwidth]{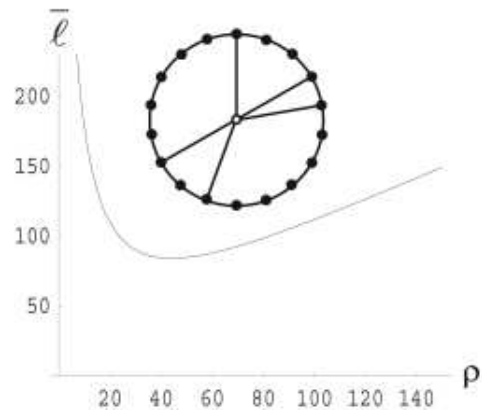}
\caption{Our model network showing transport pathways through the
central hub (connections of length $1/2$ denoted by thick lines) and around the ring (connections of
length $1$ denoted by thin lines). Graph shows average shortest path length between any two nodes in a
$n=1000$ node ring,
  with a cost-per-connection to the hub of $k=1$. There is an optimal
  value for the number of connections ($\rho\equiv p n\approx 44$) such that the average
  shortest path length $\bar{\ell}$ is a minimum.  We denote this minimal
  shortest path length as $\bar{\ell}\equiv\bar{\ell}|_{\rm
  min}$.}
\label{fig:figure2}
\end{figure}

\begin{figure}[ht]
\includegraphics[width=.35\textwidth]{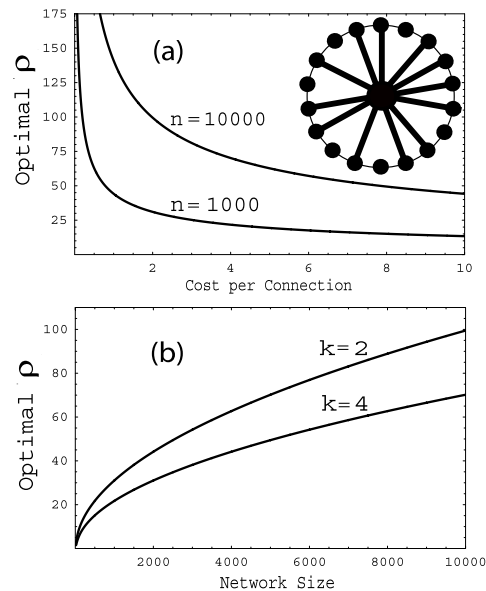}
\caption{Minimal shortest path length $\bar{\ell}|_{\rm min}$
(i.e. minimum value of $\bar{\ell}$) as obtained from Eq. (\ref{c-m-l}).  (a)
  Optimal number of connections $\rho\equiv p n$ as a function of the cost-per-connection  $k$ to the
hub.
  Results are shown for $n=1000$ and $n=10000$. (b) Optimal number
  of connections $\rho$ as a function of the network size. Results are shown for
  $k=2$ and $k=4$.}
\label{fig:figure3}
\end{figure}

\section{Results for linear and quadratic cost-functions}\label{sec:lin-quad}

For linear costs, dependent on network size and connectivity, we can show
that there exists a {\em minimum} value of the average
shortest path $\bar{\ell}$ as a function of the
connectivity to the central hub.  We denote this minimal path length
as $\bar{\ell}\equiv\bar{\ell}|_{\rm min}$.
Such a minimum is in stark contrast to the case of zero
cost per connection, where the value of $\bar{\ell}$ would just
decrease monotonically towards one with an increasing number of
connections to the hub.  The average shortest path can be
calculated from $\bar{\ell} = \sum_{\ell=1}^{n-1}\ell P(\ell)$ from which
we obtain
{
\small
  \begin{eqnarray}\label{c-m-l}
\bar{\ell }(p,n,c)=\frac{(1-p)^{n-c}\bigl(3+(n-2-c)p\bigr)}{p^2(n-1)}
\ \ \ \ \ \ \ \ \ \ \ \ \ \ \ \ \ \ \ \ \ \ \ \nonumber\\
+\frac{p\bigl(2-2c+2n-(c-1)(c-n)p\bigr)-3}{p^2(n-1)}+\frac{c(c-1)}{2(n-1)}\ \ \ \ \ \ 
  \end{eqnarray}
}
Figure 2 shows the functional form of $\bar{\ell}$ with a
cost of $1$ unit path-length per connection to the hub (i.e.  $c=knp=k\rho$,
with $k=1$). The optimal number of connections in order that $\bar{\ell}$ is a
minimum is approximately $44$ and depends on $n$.  The corresponding minimal
shortest path $\bar{\ell}|_{\rm min}$ is approximately $85$.
Figure 3(a) shows analytic results for the optimal number of connections which
yield the minimal shortest path $\bar{\ell}|_{\rm min}$, as a function of the
cost per connection for a fixed network size.  Figure 3(b) shows analytic
results for the optimal number of connections which yield the minimal shortest
path $\bar{\ell}|_{\rm min}$, as a function of the network size for a fixed cost
per connection to the hub.

To gain some insight into the underlying physics, we make some approximations
that will allow us to calculate the average shortest path analytically for a
given cost-function that is valid within the approximations.  We begin by noting that for large $n$, or more importantly large $n-c$, the
first term in Eq. (\ref{c-m-l}) may be written as
$(1-p)^{n-c} \rightarrow e^{-\rho}$.  With the condition that the cost for using
the hub isn't too high, the region containing the minimum shortest
path $\bar{\ell}\equiv\bar{\ell}|_{\rm min}$ will be at
sufficiently high $\rho$ to ignore this term, yielding a simplified form for the
average shortest path:
{\small
\begin{equation}\label{cost min lbar approx} \bar{\ell}\approx
\frac{p\bigl[2-2c+2n-(c-1)(c-n)p\bigr]-3}{p^2(n-1)}+\frac{c(c-1)}{2(n-1)}.
\end{equation}}
We may then proceed by considering that, for a fixed network size and a cost that depends on connectivity,
to locate the minima we differentiate Eq. (\ref{cost min lbar approx}) with respect to
$p$ and set the result equal to zero and obtain
{\small 
\begin{equation}\label{optimum gen cost in p}
    -\frac{2}{p^2}(1-c+n)
    -\frac{2}{p}\deriv{c}{p} 
    - c\deriv{c}{p} 
    + (1+n)\deriv{c}{p} + \frac{6}{p^3}
    -\frac{1}{2}\deriv{c}{p}=0.
\end{equation}}
We substitute into this expression the scaled connectivity, $\rho \equiv np$, and it 
then becomes 
{\small\begin{equation}\label{optimum gen cost in rho}
    \frac{2\rho}{n}\bigl(1-c+n\bigr)=\frac{\rho^3}{n^2}\biggl(n+\frac{1}{2}-c-\frac{2n}{\rho}\biggr)\deriv{c}{\rho}+6.
\end{equation}}
In the limit of $n \gg c$ and $\rho \gg 1$, the dominant terms on both sides of 
Eq. (\ref{optimum gen cost in rho}) are those in $n$ leaving
\begin{equation}\label{optimum general rho}
    \deriv{c}{\rho} = \frac{2n}{\rho^2}.
\end{equation}
From this expression we may obtain the location of the minimum of the average
shortest path for a given cost-function for which the approximations are valid.  
For example, in the case of linear cost $c = kn\rho$, we find that for the optimum number of
connections we have $\rho = \sqrt{\frac{2n}{k}}$.  Using $k=1$ and
$n=1000$, we obtain the value of the optimum number of connections as $44.7$, which
agrees well with the exact value calculated from Eq. (\ref{c-m-l}).  Inserting the optimum value for $\rho$ into
Eq. (\ref{cost min lbar approx}) and keeping the largest terms
we obtain $\bar{\ell} \approx \sqrt{8kn}$, which also agrees well with
the exact result.

We now consider quadratic cost-functions, $c=k \rho^2$. This could be 
a physically relevant cost-function when the cost for using the
central hub depends on the number of connected pairs created,
rather than the number of direct connections made to the hub.
Solving for the optimal number of connections using
Eq. (\ref{optimum general rho}) gives $\rho \approx
\sqrt[3]{\frac{n}{k}}$, corresponding to a minimum average
shortest path length $\bar{\ell} \approx \sqrt[3]{27kn^2}$.  One is also
able to consider a cost dependant on a general exponent $c=k
\rho^{\alpha}$.  This gives for the optimal number of connections
\begin{equation}
\rho \approx \biggl(\frac{2n}{\alpha k} \biggr
)^{\frac{1}{1+\alpha}}.
\end{equation}
The corresponding average shortest path is a more
complicated expression, but it scales with $k$ and $n$ as
\begin{equation}
\bar{\ell}|_{\rm{min}} \propto k^{\frac{1}{1+\alpha}}
n^{\frac{\alpha}{1+\alpha}} \ .
\end{equation}

This analysis can be adapted to the `undirected' model by using the
usual scaling relation between the models that was described above.  
For the case of linear costs on an undirected network one gets an optimal 
number of connections at $\rho \approx \sqrt{\frac{n}{k}}$ and a minimum average
shortest path of $\bar{\ell}|_{\rm{min}} \approx \sqrt{4kn}$.

\begin{figure}[ht]
\includegraphics[width=.48\textwidth]{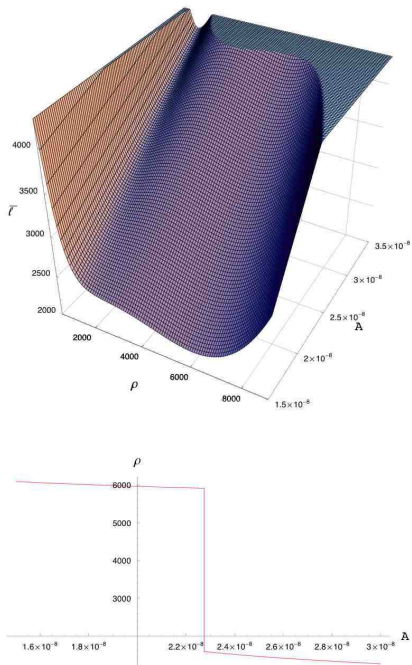}
\caption{Top: Landscape of the average shortest path length
  $\bar{\ell}$ (vertical axis) as a function of the cubic cost-function parameter $A$ and the
    average number of connections to the central hub $\rho$.  Bottom: The value of
    $\rho$ corresponding to a global minima in $\bar{\ell}$, as a function of the
    cubic cost-parameter $A$.}
\label{fig:figure4}
\end{figure}

\section{Results for non-linear cost-functions}\label{sec:cubic}
We now consider the functional form of $\bar{\ell}$ for non-linear cost 
functions, specifically a cubic cost-function and a `step' cost-function.
We show that for these non-linear cost-functions, a novel and highly non-trivial 
phase-transition can arise.
First we consider the
case of a general cubic cost-function: 
\begin{equation} c(\rho) = A\rho^3 +
  B\rho^2 + C\rho + D, 
\end{equation} 
where $\rho$ is the scaled probability, $\rho = pn$ and $A, B, C,
D\in\mathbf{R}$. In order to demonstrate the richness of the phase transition
and yet still keep a physically reasonable model, we choose the minimum in this
cubic function to be a stationary point.  Hence the cost-function remains a
monotonically increasing function, but features a regime of intermediate
connectivities over which congestion costs remain essentially flat (like the
`fixed charge' for London's congestion zone).  Since we are primarily concerned
with an optimization problem, we can set the constant $D=0$. Hence 
\begin{equation}\label{cubic-cost}
    c(\rho) = A\rho^3 - 3Ar\rho^2 + 3Ar^2\rho, 
\end{equation} 
where $r=\frac{-B}{3A}$ is the location of the stationary point.  Substituting
into Eq. (\ref{c-m-l}) yields the shortest path distribution for this particular
cost-function in terms of the parameters $A$, $r$, $p$ and $n$.  The result is
too cumbersome to give explicitly -- however we emphasize that it is
straightforward to obtain, it is exact, and it allows various limiting cases to
be analyzed analytically. 

Figure 4 (top) shows the value of the average shortest path $\bar{\ell}$ for
varying values of $\rho$ and $A$.  As can be seen, the optimal network structure
(i.e. the network whose connectivity $\rho$ is such that $\bar{\ell}$ is a
global minimum) changes abruptly from a high connectivity structure to a low
connectivity one, as the cost-function parameter $A$ increases.  Figure 4
(bottom) shows that this transition resembles a first-order phase transition. At
the transition point $A=A_{\rm crit}$, both the high and low connectivity
structures are optimal. Hence there are two structurally inequivalent networks
having identical (and optimal) functional properties. As we move below or above
the transition point (i.e. $A<A_{\rm crit}$ or $A>A_{\rm crit}$ respectively)
the high or low connectivity structure becomes increasingly superior. 

Using the same approximations as those in Sec. \ref{sec:lin-quad}, we
may estimate the approximate value of the cubic function parameter, $A$, that will
lead to two minima in the average shortest path distribution. We proceed by solving 
Eq. (\ref{optimum general rho}) with our cubic function as the cost-function:
\begin{equation} \label{cubic double condition}
\rho^2 \deriv{c}{\rho} - 2n =3A(\rho^2-2r\rho+r^2)\rho^2 - 2n =0 \
.
\end{equation}
The solutions to this equation are then stationary points in $\bar{\ell}$, 
and at least three stationary points are required for the distribution to have
multiple minima.  We thus have $\rho=0$ and $\rho=(3\pm\sqrt{7})r$. Inserting the
central value, $(3-\sqrt{7})r$, into Eq. (\ref{cubic double condition}) gives an 
approximate lower bound for $A$:
\begin{equation} \label{cubic A}
A_{\rm{min}} \sim \frac{n}{r^4} \ .
\end{equation}
Although this calculation does not give us the value of $A_{\rm{crit}}$, 
it is expected (and results confirm such a conjecture) to be close to
$A_{\rm{min}}$.  From this analysis, we can also see that both the location of
the minima and the distance between them is governed by the cubic parameter $r$.

We have checked that similar structural transitions can arise for higher-order
nonlinear cost-functions. In particular we  demonstrate here the extreme case of
a `step' function, where the cost is fixed until the connectivity to the central
hub portion reaches a particular threshold value. As an illustration, we
consider the particular case: 
\begin{equation}\label{step-cost}
c(\rho,r_0) = 50 \bigg(\sum_{i=1}^{50} {\rm Sgn} (\rho - i r_0) + 50\bigg),
\end{equation}
where ${\rm Sgn}(x)=-1,0,1$ depending on whether $x$ is negative, zero, or
positive respectively, and $r_0$ determines the frequency of the jump in the
cost.  Figure 5 (top) shows the average shortest path $\bar{\ell}$ for this step
cost-function (Fig. 5 (bottom)) as $\rho$ and $r_0$ are varied.  A multitude of
structurally-distinct yet optimal network configurations emerge.  As $r_0$
decreases, the step-size in the cost-function decreases and the cost-function
itself begins to take on a  linear form -- accordingly, the behavior of
$\bar{\ell}$ tends towards that of a linear cost model with a single
identifiable minimum.  Most importantly, we can see that once again a
gradual change in the cost parameter leads to an abrupt change in the structure
of the optimal (i.e. minimum $\bar{\ell}$) network.

\begin{figure}[ht]
\includegraphics[width=.48\textwidth]{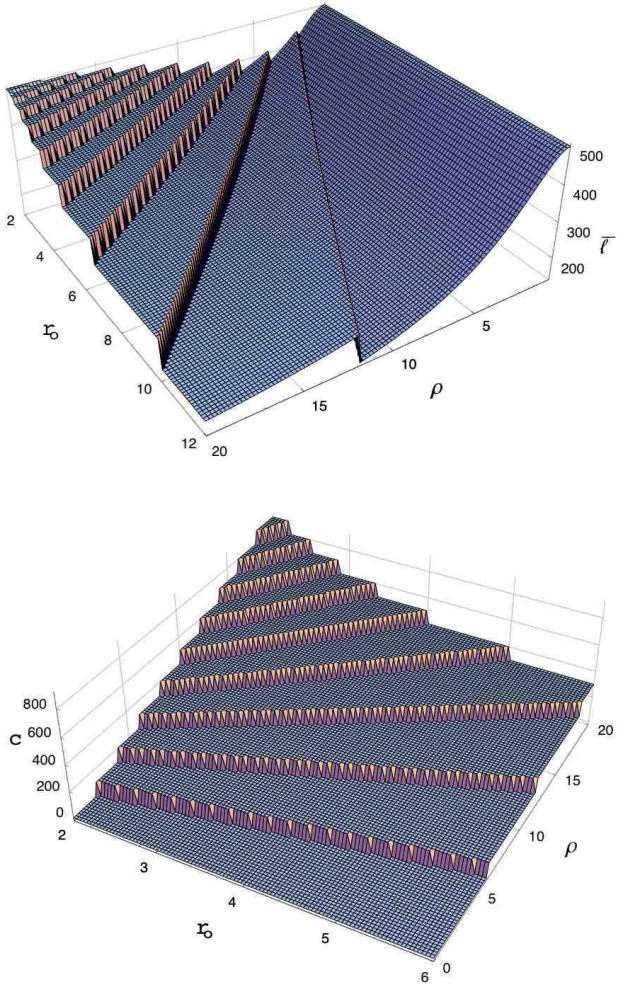}
\caption{ Top: Landscape of the average shortest path length
  $\bar{\ell}$ (vertical axis) as a function of the
`step' cost-function parameter $r_0$ and the
    connectivity $\rho$. Bottom:
The `step' cost-function as a function of the step-frequency parameter $r_0$ and $\rho$.  As $r_0$ decreases, the cost-function becomes increasingly 
linear. } \label{fig:figure5}
\end{figure}

\section{The ring-hub structure as a network motif}\label{sec:motif}
We have allowed our ring-and-hub networks to seek optimality by modifying their
radial connectivity while maintaining a single ring. Relaxing this constraint to
allow for transitions to multiple-ring structures yields a number of related findings.  
In particular, allowing both the radial connectivity and the number of rings to
change yields abrupt transitions between optimal networks with different radial
connectivities {\em and} different numbers of rings.  One could, for example, consider 
this to be a model of a complicated fungal structure (Figs. 1(b), 1(e)) or of the
interactions between hierarchies within an organization, such as in Fig. 6.

\begin{figure}[ht]
\includegraphics[width=.5\textwidth]{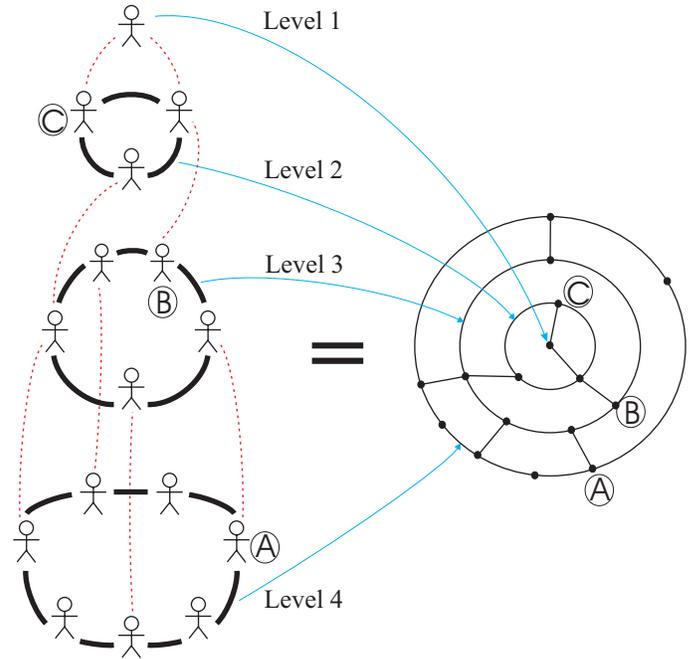}
\caption{Schematic description of hierarchies in a human organization, institution or company. As shown, this diagram can be re-drawn as a multiple-ring-and-hub structure. Similar networks are likely to exist in a range of social systems. } \label{fig:figure6}
\end{figure}

To analyze analytically
such multiple-ring structures we introduce the following renormalization scheme.
Consider the two-ring-and-hub network in Fig. 1(f). For paths which pass near
the center, there is a contribution to the path-length resulting from the fact
that the inner ring has a network structure which needs to be traversed. Hence
the inner-ring-plus-hub portion acts as a {\em renormalized hub} for the outer
ring. In short, the ring-plus-hub of Eq. (1) can be treated as a `cost motif'
for solving multiple-ring-and-hub problems, by allowing us to write a recurrence
relation which relates the average shortest path in a network with $i+1$ rings
to that for $i$ rings:
{
\small
  \begin{eqnarray}\label{c-m-l-rings}
\bar{\ell }_{i+1}(p_{i+1},n_{i+1},c)= 
\ \ \ \ \ \ \ \ \ \ \ \ \ \ \ \ \ \ \ \ \ \ \ \ \ \ \ \ \ \ \ \ \ \ \ \ \ \ \ \
\ \ \ \ \ \ \ \ \ \ \ \ \ \ \ \ \ \ \ \ \ \ \ \ \ \ \ \ \ \ \ \ \ \ \ \ \ \ \ \ 
\ \ \ \ \ \ \ \ \ \ \ \ \ \ \ \ \ \ \ \ \ \ \ \ \ \ \ \ \ \ \ \ \ \ \ \ \ \ \nonumber\\
\frac{(1-p_{i+1})^{n_{i+1}-\bar{\ell }_{i}(p_{i},n_{i},c)}\bigl(3+(n_{i+1}-2-\bar{\ell }_{i}(p_{i},n_{i},c))p_{i+1}\bigr)}{p_{i+1}^2(n_{i+1}-1)}
\ \ \ \ \ \ \ \ \ \ \ \ \ \ \ \ \ \ \ \ \ \ \ \ \ \ \ \ \ \ \ \ \ \ \ \ \ \ \ \
\ \ \ \ \ \ \ \ \ \ \ \ \ \ \ \ \ \ \ \ \ \ \ \ \ \ \ \ \ \ \ \ \ \ \ \ \ \ \ \
\ \ \nonumber\\
+\frac{2p_{i+1}\bigl(1-\bar{\ell }_{i}(p_{i},n_{i},c)+n_{i+1})}{p_{i+1}^2(n_{i+1}-1)}
\ \ \ \ \ \ \ \ \ \ \ \ \ \ \ \ \ \ \ \ \ \ \ \ \ \ \ \ \ \ \ \ \ \ \ \ \ \ \ \
\ \ \ \ \ \ \ \ \ \ \ \ \ \ \ \ \ \ \ \ \ \ \ \ \ \ \ \ \ \ \ \ \ \ \ \ \ \ \ \
\ \ \ \ \ \ \ \ \ \ \ \ \ \ \ \nonumber\\
-\frac{p_{i+1}\bigl((\bar{\ell }_{i}(p_{i},n_{i},c)-1)(\bar{\ell }_{i}(p_{i},n_{i},c)-n_{i+1})p_{i+1}\bigr)-3}{p_{i+1}^2(n_{i+1}-1)}
\ \ \ \ \ \ \ \ \ \ \ \ \ \ \ \ \ \ \ \ \ \ \ \ \ \ \ \ \ \ \ \ \ \ \ \ \ \ \ \
\ \ \ \ \ \ \ \ \ \ \ \ \ \ \ \ \ \ \ \ \ \ \ \ \ \ \ \ \ \ \ \ \ \ \ \ \ \ \ \
\ \ \ \ \ \ \ \ \ \ \ \ \ \ \ \nonumber\\
+\frac{\bar{\ell }_{i}(p_{i},n_{i},c)(\bar{\ell }_{i}(p_{i},n_{i},c)-1)}{2(n_{i+1}-1)}\, \
\ \ \ \ \ \ \ \ \ \ \ \ \ \ \ \ \ \ \ \ \ \ \ \ \ \ \ \ \ \ \ \ \ \ \ \ \ \ \ \
\ \ \ \ \ \ \ \ \ \ \ \ \ \ \ \ \ \ \ \ \ \  \ \ \ \
\ \ \ \ \ \ \ \ \ \ \ \ \ \ \ \ \ \ \ \ \ \ \ \ \ \ \ \ \ \ \ \ \ \ \ \ \ \ 
\ \ \ \ \ \ \ \ \ \ \ \ \ 
\end{eqnarray}
}
where $i \geq 0$ and $\bar{\ell}_{0} = c$ with $c$ being a general cost for
the inner-most hub. The case $i=0$ is identical to Eq. (1). As before, $p_{i+1}$
represents the probability of a link between rings $i+1$ and $i$ and $n_{i+1}$
is the number of nodes in ring $i+1$.  

We investigate the properties of our renormalized $N$-ring network by placing
a number of constraints on the parameters and observing the average shortest
path behaviour, so we may once again determine regimes of functionally optimal
network configurations.
We begin by increasing the number of rings to $N=2$, with the constraint that the 
number of nodes on each ring is fixed.  We find that the configuration that yields 
the minimum average shortest path length has all the ring probabilities, $p_i$, 
equal.  Figure 7 demonstrates the average shortest path distribution for such a 
case.  Figure 7 also demonstrates the accuracy of our analytic renormalized result, as compared to a full-scale numerical calculation for $\bar{\ell}$. If we allow the number of nodes on the two rings to be different, we find
that the configuration that optimizes the shortest path favours a larger number
of connections on the ring with the most nodes.
Returning to the original configuration, $N=2$ with an equivalent number of
nodes on each ring, we consider the effect of a cost on the central hub of the
inside ring.  We find that a greater number of connections on the ring without 
costs optimizes the network, as one might expect.

\begin{figure}[ht]
\includegraphics[angle=-90,width=.48\textwidth]{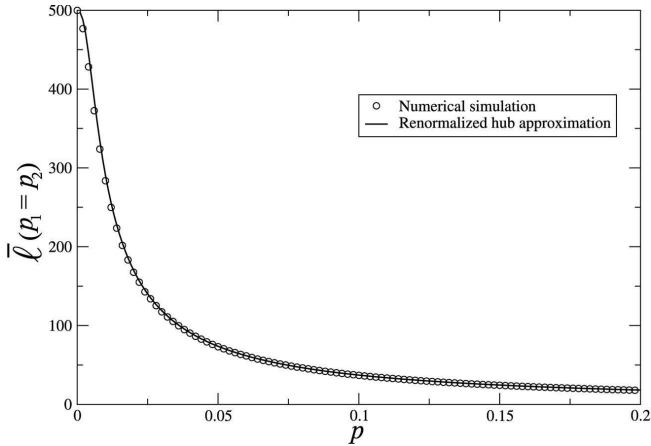}
\caption{Average shortest path $\bar{\ell}$ for a network with $N=2$ rings, each of size $n=10^3$ 
and with equal connectivity $p_1=p_2\equiv p$.} \label{fig:figure7}
\end{figure}

The addition of further rings, such that $N>2$, leads to some interesting
results.  For a network with an equivalent number of nodes on each ring the
optimal configuration remains such that all the ring probabilities, $p_i$,
are equal.  However, for $N\gg2$, we begin to see a deviation: connections should 
be moved to the outer rings of the network (those furthest away from the hub) in
increasing numbers out to the edge of the network to obtain the minimum shortest
path.  We thus consider the properties of a network with varying $N$; equal ring
probabilities $p_i$; inner ring costs, $c_0$, that vary from $0 \to \infty$; both fixed 
numbers of nodes on each ring, $n_i = n =$ constant, and varying numbers of nodes on 
each ring such that $n_i \varpropto i$.  Figure 8 shows the average shortest path
distribution for several such cases.  Interestingly, for large $N$ the system
becomes indifferent to the cost of the central hub, $c_0$, and all the
distributions converge, in both the case of fixed and varying numbers of nodes
on each individual ring.  This is suggestive of an effective `skin effect', as the
center of the network becomes effectively disconnected after the addition of a large
number of rings.  The redundance of the central portion of the network offers an
explanation for an earlier finding: that as we approach $N\gg2$ the optimal
configuration ceases to be such that all innermost ring probabilities, $p_i$, are 
equal.  We find that to optimize the network we need to move the connections
into the skin, more than likely as a result of the detachment of the center of the
network from the whole.

By comparison of the shortest path values of our multiple ring networks,
we have found a further important result.  We find that there are 
optimal network structures with different numbers of rings and radial 
connectivities, yet which have the {\em same} average shortest path
length across them \cite{note}. Hence, as before, optimal network structures
exist which are structurally very different, yet functionally equivalent. Figure
9 shows an explicit example of two such functionally equivalent, optimal
networks. It is remarkable that these images are so similar to the real fungi
shown in Figs. 1(a) and (b).

\section{Conclusion}
In summary, we have analyzed the interplay between the structure and function within a class of biologically-motivated networks, and have uncovered a novel structural phase transition. Depending on the system of interest (e.g.
fungus, or road networks) these transitions between inequivalent structures
might be explored in real-time by adaptive re-wiring, or over successive
generations through evolutionary forces or `experience'.  Through the use of an
approximation, we treated the original network as a cost-motif for a 
larger network and considered the circumstances under which such networks obtained
their optimal functional structure.  The equivalence in function, defined by the
networks transport properties, between various topologically distinct structures
may provide insight into the existence of such disparate structure in real
fungi.
An important further implication of this work is that in addition to searching 
for a universality in terms of network structure, one might fruitfully consider 
seeking universality in terms of network {\em function}.

\section*{Acknowledgements}
We kindly acknowledge L. Boddy, J. Wells, M. Harris and G. Tordoff (University
of Cardiff) for the fungal images in Figs. 1(a), (b) and (e). N.J. and M.F.
acknowledge the support of EU through the  MMCOMNET project, and N.J. acknowledges the support of the EPSRC through the Life Sciences Program.

\begin{figure}[ht]
\includegraphics[angle=-90,width=.48\textwidth]{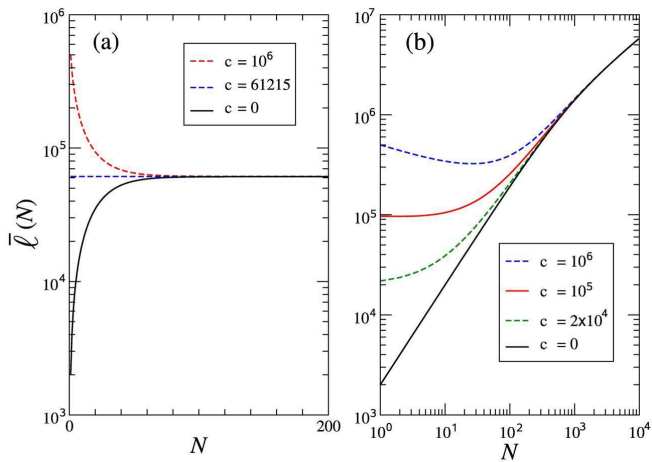}
\caption{Average shortest path $\bar{\ell}$ as a function of the number of rings $N$ that make up the
  network. In (a) all rings are of size $n=10^6$ and in (b) the rings increase
  according to $n_i = 10^6 i$ where $i$ is the ring index. In both cases, the probability of connectivity of a node is a constant and equal to $p_i=0.001$.  The innermost
  ring has a hub with a constant cost $c_0$.  Here $c_0$ ranges from $0$ to effectively
  infinite. In all cases the limiting value for $\bar{\ell}$ is the
  same, demonstrating the `skin effect' and hence the effective disconnection of the inner rings.}  \label{fig:figure8}
\end{figure}

\begin{figure}[ht]
\includegraphics[width=.5\textwidth]{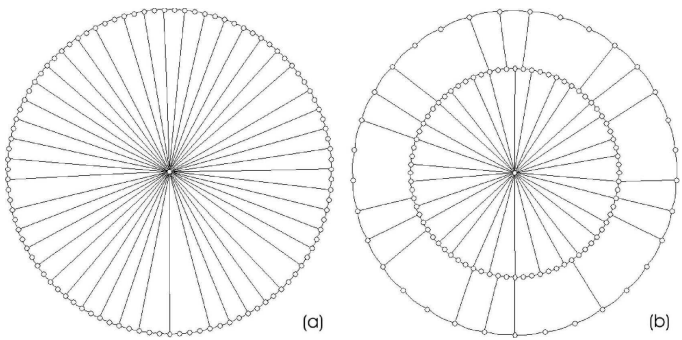}
\caption{Two structurally inequivalent networks, which are functionally {\em equivalent} in the sense that they have the same average shortest path across them. The average shortest path across the structure is
    the same for both networks, and both networks are themselves optimal (i.e. minimum $\bar{\ell}$). Nutrients found on the perimeter of each structure should therefore take the same average time to cross it -- this in turn implies that both structures would co-exist in the same favourable environmental conditions. 
        (a) A single-ring-and-hub network with a linear cost-function.   (b) A two-ring-and-hub configuration. The inner
    ring-hub structure has the same cost-function as in (a). The similarity to the real fungi in Figs. 1(a) and (b) is striking.} \label{fig:figure9}
\end{figure}

\end{document}